\documentclass[twocolumn,aps,prd,showpacs,amsmath,amssymb,nofootinbib,nobibnotes]{revtex4-2}
\usepackage{anyfontsize}
\usepackage{amssymb}
\usepackage{amsmath}
\usepackage{amsfonts}
\usepackage{amsbsy}
\usepackage[utf8]{inputenc} 
\usepackage{newtxtext}
\usepackage{newtxmath}

\usepackage[normalem]{ulem}

\usepackage{tensor}
\usepackage{mathrsfs}

\usepackage{bm}

\usepackage{graphicx}
\usepackage{epsfig}
\usepackage{epstopdf}

\usepackage[usenames,dvipsnames]{xcolor}

\usepackage{verbatim,mathtools,needspace,enumitem,etoolbox}
\usepackage{physics}

\definecolor{linkcolor}{rgb}{0.0,0.3,0.5}

\usepackage[unicode, colorlinks=true, linkcolor=linkcolor, citecolor=linkcolor, filecolor=linkcolor,urlcolor=linkcolor, pdfusetitle]{hyperref}

\usepackage{epstopdf}

\usepackage{bm}
\usepackage{dcolumn}

\usepackage{longtable}
\usepackage{enumerate}
\usepackage[normalem]{ulem}

\hypersetup{colorlinks=true,
citecolor=blue,
linkcolor=blue,
urlcolor=blue}
\usepackage{subfigure}
\usepackage{amsmath}
\usepackage{float}
\usepackage{mathrsfs}
\usepackage[dvipsnames]{xcolor}
\usepackage{soul}

\definecolor{darkgreen}{RGB}{1,212,57}

\usepackage{multirow}

\usepackage{graphicx}
\usepackage{dcolumn}
\usepackage{bm}

\begin{document}


\title{Quasinormal modes of a holonomy corrected Schwarzschild black hole}

\author{Zeus S. Moreira}
\email{zeus.moreira@icen.ufpa.br}
\affiliation{%
	Programa de Pós-Graduação em Física, Universidade Federal do Pará, 66075-110, Belém, Par{\'a}, Brazil 
}%
\affiliation{Departamento de Matem\'atica da Universidade de Aveiro and Centre for Research and Development  in Mathematics and Applications (CIDMA), Campus de Santiago, 3810-183 Aveiro, Portugal.}
\author{Haroldo C. D. Lima Junior}%
\email{haroldolima@ufpa.br}
\affiliation{%
	Programa de Pós-Graduação em Física, Universidade Federal do Pará, 66075-110, Belém, Par{\'a}, Brazil 
}%
\affiliation{Departamento de Matem\'atica da Universidade de Aveiro and Centre for Research and Development  in Mathematics and Applications (CIDMA), Campus de Santiago, 3810-183 Aveiro, Portugal.}

\author{Luís C. B. Crispino}
\email{crispino@ufpa.br}
\affiliation{%
	Programa de Pós-Graduação em Física, Universidade Federal do Pará, 66075-110, Belém, Par{\'a}, Brazil 
}%
\affiliation{Departamento de Matem\'atica da Universidade de Aveiro and Centre for Research and Development  in Mathematics and Applications (CIDMA), Campus de Santiago, 3810-183 Aveiro, Portugal.}

\author{Carlos A. R. Herdeiro}
\email{herdeiro@ua.pt}
\affiliation{Departamento de Matem\'atica da Universidade de Aveiro and Centre for Research and Development  in Mathematics and Applications (CIDMA), Campus de Santiago, 3810-183 Aveiro, Portugal.}


\begin{abstract}
	We analyze the quasinormal modes (QNMs) of a recently obtained solution of a Schwarzschild black hole (BH) with corrections motivated by Loop Quantum Gravity (LQG). This spacetime is regular everywhere and presents the global structure of a wormhole, with a minimal surface whose radius depends on a LQG parameter. We focus on the investigation of massless scalar field perturbations over the spacetime. We compute the QNMs with the 
WKB approximation, as well as the continued fraction method. The QNM frequency orbits, for $l=0$ and $n>0$, where $l$ and $n$ are the multipole and overtone numbers, respectively, are self-intersecting, spiraling curves in the complex plane. These orbits accumulate to a fixed complex value corresponding to the QNMs of the extremal case. 
We obtain that, for small values of the LQG parameter, the overall damping decreases as we increase the LQG parameter. Moreover the spectrum of the quantum corrected black hole exhibits an oscillatory pattern, which might imply in the existence of QNMs with vanishing real part. This pattern suggests that the limit $n\rightarrow \infty$ for the real part of the QNMs is not well-defined, what differs from Schwarzschild's case. We also analyze the time-domain profiles for the scalar perturbations, showing that the LQG correction does not alter the Schwarzschild power-law tail. We compute the fundamental mode from the time profile by means of the Prony method, obtaining excellent agreement with the two previously mentioned methods.

\end{abstract}

\maketitle



\section{Introduction}
\label{Sec. I}


The detection of gravitational waves (GWs) marks the beginning of GW astronomy \cite{Ligo_a,Ligo_b} and creates great expectations for the future of gravitational physics research. Any orbiting pair of astrophysical objects produces GWs, but only those sufficiently compact and moving very rapidly can produce detectable signals for the current generation of GWs detectors. This makes BH binaries ideal systems for detecting GWs. The collision of BHs can be divided in three stages: \textit{(i) inspiral:} The BHs orbit around each other, getting closer due to loss of energy through GWs; \textit{(ii) merger:} the actual collision of the two BHs and \textit{(iii) ringdown:} The merged BH relaxes to its equilibrium form (widely  believed to be a Kerr BH \cite{Kerr_1963}). The GW signal produced by the binary carries a very  characteristic signature \cite{Pretorious_2005,Campanelli_2006,Baker_2006}, which in turn can reveal properties of the BH itself \cite{Echeverria_1988,Finn_1992,Berti_2006}.  

In a perturbed physical system, the modes of vibration associated with energy dissipation are called quasinormal modes (QNMs). Thus, the ringdown phase of the coalescence of two BHs is essentially characterized by the corresponding QNMs \cite{Berti_2009,Kokkotas_1999}. The study of BH perturbations began with the work of Regge and Wheeler \cite{Regge_1957} and was further developed by Zerilli \cite{Zerilli_1970a,Zerilli_1970b}. The problem of finding the QNM frequencies was investigated for the first time in a famous paper written by Chandrasekhar and Detweiler as a non self-adjoint boundary problem \cite{Chandrasekhar_1975}. Thus, we loose the nice properties of self-adjoint problems, such as completeness and normalizability of the eigenfunctions and the spectrum becomes complex \cite{Lemos,Konishi}. The real part of the eigenfrequencies are the standard oscillation frequency, whereas the imaginary part is related to the wave damping. 

Starting in the 1960s, due to several astronomical discoveries related with pulsar, quasars and cosmic background radiation, Einstein's theory of GR experienced a new series of experimental confirmations \cite{dinverno,Will}. More recently, in addition to the already mentioned detection of GWs, the shadow images of M87* and Sgr A* were obtained~\cite{M87,Sgr}. GR had its birth in the beginning of the 20st century and now it is  enjoying a more mature and robust era, both theoretically and experimentally.

Despite these previous achievements, most relativists believe that GR cannot be the final theory of gravity and should be replaced by some quantum theory. The very early universe \cite{Hawking_1970}, the interior of BHs \cite{Penrose_1965} and the last stages of BH evaporation \cite{Hawking_1975} are examples of physical scenarios where quantum effects play a fundamental role and GR no longer gives a precise description of the gravitational field. This is one of the biggest open questions in theoretical physics to date, i.e. how to reconcile gravity with quantum mechanics.

The idea to canonically quantize gravity considering as canonical variables the spatial metric and its conjugated momentum, led to some problems~\cite{DeWitt_1967}. Since the constraints equations are non-polynomial functions of the canonical variables, their corresponding operator equations in the quantum formulation are not well-defined \cite{Thiemann_2001}. Due to Sen, Ashtekar and Barbero, a new set of coordinates was found, the Ashtekar-Barbero connection variables, such that the constraints equations were reduced to polynomial expressions \cite{Sen_1982,Ashtekar_1986,Barbero_1995}. By writing GR in terms of the Ashtekar-Barbero variables, it is possible to put the theory in a framework very similar to other quantum field theories, where quantization techniques have already been developed \cite{Wilson_1974}. The early construction of Loop Quantum Gravity (LQG) was based on the quantization of GR, in terms of a smeared version of the connection variables in a background independent fashion. 
As some of the important results of LQG, we can mention
the construction of singular-free cosmological models \cite{Bojowald_2001}, the quantization of spherically symmetric vacuum spacetime \cite{Gambini_2013}, as well as the derivation of the Hawking-Bekenstein entropy \cite{Ashtekar_1998}.

Working within the full machinery of LQG is very challenging
 and some effective models have shown to be useful in understanding how quantum gravity effects might look like.
There are several works applying modifications to GR for cosmological models \cite{Bojowald_review,Ashtekar_2006} and also for spherically symmetric spacetimes, such as Schwarzschild \cite{Vakili_2018,Bodendorfer_2019,Achour_2018} and Reissner-Nordström solutions \cite{Tibrewala_2012}.

In Refs. \cite{Bardaji_2022a,Bardaji_2022b} an effective spherically symmetric spacetime is proposed, which is not singular and presents a global structure of a wormhole whose minimal surface is hidden by an event horizon. Here we calculate the scalar QNMs of this quantum corrected BH, investigating how its spectrum deviates from the well-known Schwarzschild case.

The remaining of this paper is organized as follows. In Sec. \ref{Sec. II} we review some aspects of the solution obtained in Refs. \cite{Bardaji_2022a,Bardaji_2022b}, highlighting its main properties. In Sec. \ref{Sec. III} we investigate the dynamics of a massless scalar field over the quantum corrected spacetime and review the corresponding boundary problem of QNMs. In Sec. \ref{Sec. IV} we revisit two methods for calculating QNM frequencies, namely the third order Wentzel-Kramers-Brillouin (WKB) approximation, as well as the Leaver's continued fraction method.  In Sec. \ref{Sec. V} we exhibit a selection of our numerical results. We first compare, as a consistency check, the third order WKB results and the ones obtained via continued fraction calculations. We also compute, with the Leaver method, 
the first 30 overtones for the modes $l=0$ and $l=1$. 
We present our final remarks in Sec. \ref{Sec. VI}. We use natural units, such that $c=G=\hbar=1$.

\section{Effective quantum corrected Schwarzschild spacetime}
\label{Sec. II}

The authors of Ref.~\cite{Bardaji_2022a,Bardaji_2022b} reported the following line element:
 \begin{equation}\label{metric}
	ds^2=- f(r)dt^2+\left[\left(1-\frac{r_0}{r}\right)f(r)\right]^{-1}dr^2+r^2d\Omega^2,
\end{equation}
where $r_0<2M$ is a LQG parameter, $f(r)\equiv 1-2M/r$ and $d\Omega^2$ is the line element of the 2-sphere. 
This metric represents a static, spherically symmetric and asymptotically flat spacetime. The horizon is located at the hypersurface $r=r_h=2M$, similarly to what we have in Schwarzschild spacetime. Nonetheless, the quantity $M$ cannot be simply interpreted as the mass of the BH. As pointed out in Ref. \cite{Bardaji_2022b}, the different geometric definitions of mass, namely, the Komar, ADM and Misner-Sharp masses, need to be taken into account. These quantities are given by 
\begin{subequations}
	\begin{align}
	\label{Komar}
		&M_{\text{K}}=M\sqrt{1-\frac{r_0}{r}},\\
		\label{ADM}
		&M_{\text{ADM}}=M+\frac{r_0}{2},\\
		\label{MS}&M_{\text{MS}}=M+\frac{r_0}{2}-\frac{M r_0}{r},
	\end{align}
\end{subequations}
where $M_{\text{K}},\ M_{\text{ADM}}$ and $M_{\text{MS}}$ are the Komar, ADM and Misner-Sharp mass, respectively. The Komar and Misner-Sharp masses do not need to coincide, since the quantum corrected spacetime is not a solution of the Einstein's equations \cite{Beig_1976}. However, in the limit that $r$ goes to infinity, for spherically symmetric and asymptotically flat spacetimes, the ADM and Misner-Sharp masses must be equal \cite{Hayward_1996}, what is indeed the case. The BH parameters, $M$ and $r_0$ can be redefined in a geometric invariant way, according to 
\begin{subequations}
		\begin{align}
	&M=\lim\limits_{r\rightarrow \infty}M_{\text{K}},\\
	&r_0=2\lim\limits_{r\rightarrow \infty}(M_{\text{MS}}-M_{\text{K}}).
	\end{align}
\end{subequations}

\begin{figure}
	\centering
	\includegraphics[width=7cm]{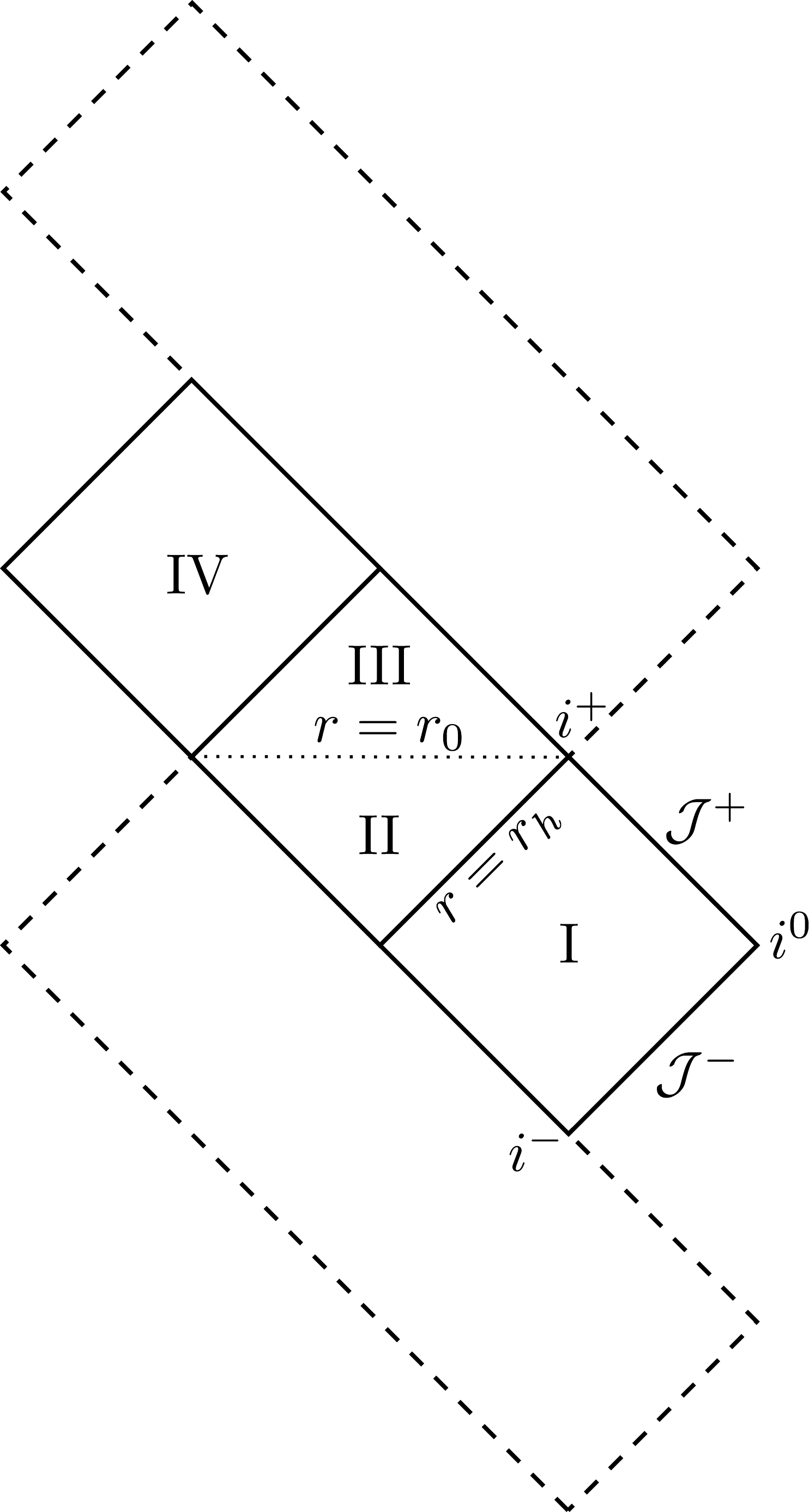}
	\caption{Penrose diagram representing the global structure of the spacetime proposed in Refs. \cite{Bardaji_2022a,Bardaji_2022b}, which corresponds to a wormhole solution. The hypersurface $r=r_0$ represents a transition surface between the BH and the white hole regions.}
	\label{fig1}
\end{figure}

In FIG.\ref{fig1} we display the Penrose diagram of the spacetime maximal extension \cite{Bardaji_2022b}. Region I stands for the asymptotically flat region in which $r\in(r_h,\infty)$. This patch have the usual conformal infinities, namely, the timelike infinities, $i^{-}$ and $i^{+}$, the null infinities,  $\mathcal{J}^{-}$ and $\mathcal{J}^{+}$ and the spatial infinity $i^0$. Region II stands for the BH region and corresponds to $r\in(r_0,r_h)$. The remaining regions III and IV, which cannot be covered by the coordinate system $(t,r,\theta,\varphi)$, are the white hole region and another asymptotically flat region, respectively. The bottom and upper regions in blank with dashed contour are  copies of the middle structure  \cite{Bardaji_2022b}.

This spacetime has a global structure of a wormhole with minimal surface area $4\pi r_0^2$ (see FIG. \ref{fig1}). The radius $r_0$ defines the minimal spacelike hypersurface separating the trapped BH interior from the anti-trapped WH region. Therefore, the effective quantum Schwarzschild spacetime, differently from the Schwarzschild BH, is regular everywhere, as can be verified by computing the curvature scalar of this spacetime.

\section{Scalar perturbations}
\label{Sec. III}

The dynamics of a massless scalar field $\Phi$ is determined by the Klein-Gordon equation
\begin{equation}\label{kg}
	\nabla_\mu\nabla^\mu\Phi=\frac{1}{\sqrt{-g}}\partial_\mu \left(\sqrt{-g}g^{\mu \nu}\partial_\nu \Phi\right)=0,
\end{equation}
where $g$ is the metric determinant and $g^{\mu\nu}$ are the contravariant components of the metric tensor. Due to spherical and time translation symmetries, the scalar field admits the product decomposition given by
\begin{equation}\label{field}
	\Phi(x^\mu)=\frac{\psi_{\omega l}(r)}{r}Y_{lm}(\theta,\varphi)e^{-i\omega t},
\end{equation}
where $Y_{lm}(\theta,\varphi)$ are the spherical harmonics.

\begin{figure}
	\centering
	\includegraphics[width=\columnwidth]{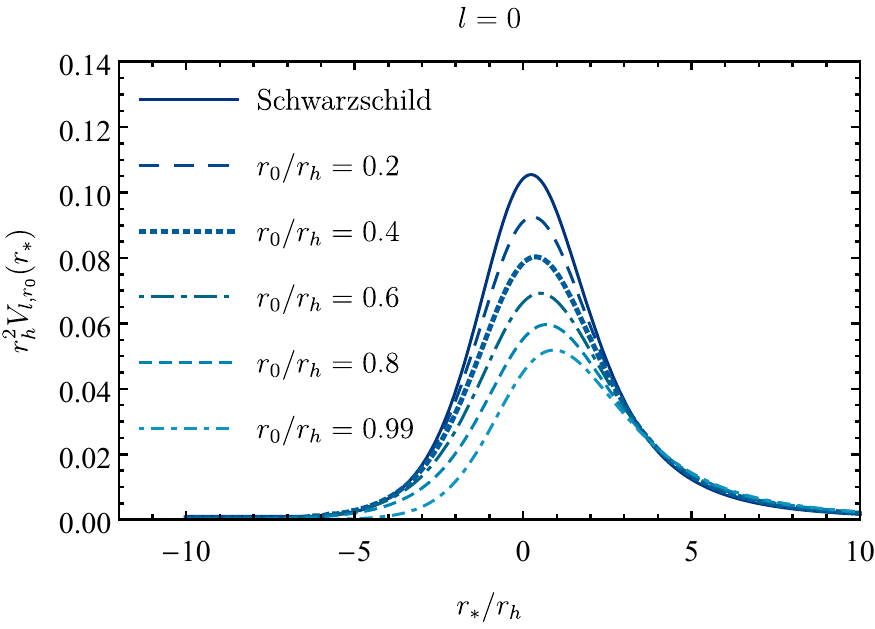}
	\caption{Effective potential given in Eq. (\ref{potential}) as a function of the tortoise coordinate for various values of $r_0$ and $l=0$.}
	\label{fig2}
\end{figure}

Inserting the metric components given in Eq. (\ref{metric}), as well as the field decomposition given in Eq. (\ref{field}), into Eq. (\ref{kg}), we obtain a Schrödinger-like equation for the radial part, which is given by
\begin{equation}\label{radial}
	\frac{d^2 \psi_{\omega l}}{dr_*^2}+\left(\omega^2 -V_{l,r_0}[r(r_*)]\right)\psi_{\omega l}=0,
\end{equation}
where the effective potential $V_{l,r_0}(r)$ is defined by
\begin{equation}\label{potential}
	V_{l,r_0}(r)=f(r) \left(\frac{l(l+1)}{r^2}+\frac{4M +r_0}{2 r^3}-\frac{3 M r_0}{ r^4}\right),
\end{equation}
and $r_*$ is the tortoise coordinate:
\begin{equation}
	dr_*=\frac{dr}{f(r)\sqrt{1-r_0/r}}.
\end{equation}
The effective potential is illustrated in FIG. \ref{fig2}, where we see that the maximum value of the potential decreases as we increase the LQG parameter $r_0$. 

To calculate the scalar QNMs of the loop quantum corrected Schwarzschild spacetime we have to solve Eq. (\ref{radial}), imposing the boundary conditions
\begin{equation}\label{bc}
	\psi_{\omega l}(r_*)\approx \begin{cases}
		e^{-i \omega r_*}\approx(r-2M)^{\frac{-2i\omega M}{\sqrt{1-\frac{r_0}{2M}}}} & r_*\rightarrow-\infty\\
		e^{+i \omega r_*}\approx e^{i \omega r}r^{2i\omega M+\frac{i \omega r_0}{2}} & r_*\rightarrow+\infty
	\end{cases}.
\end{equation}
Eqs. (\ref{radial}) and (\ref{bc}) define an eigenvalue problem for $\psi_{\omega l}$ with eigenvalue $\omega$ in the domain $r\in(2M,\infty)$. We expect that the spectrum is a countable infinite set $\{\omega_n|\ n=0,1,...\}$, where $n$ enumerates the eigenfrequencies in increasing imaginary part magnitude order, the so-called overtones. 

In general, an expression for the spectrum cannot be written in a closed analytical form, not even for Schwarzschild. Thus, it is common to implement approximate and numerical methods to treat the problem of QNMs. In the next session, we implement the third order WKB, continued fraction and the Prony methods to compute the scalar eigenfrequencies of the loop quantum corrected Schwarzschild spacetime.

\section{scalar QNMs calculations}
\label{Sec. IV}

\subsection{WKB approximation}
\label{Subsec. IV.A}

The first method we implement for the calculation of scalar QNMs is the third order WKB approximation. The WKB method is a semianalytic technique, first applied to BH scattering problems by Schutz and Will \cite{Schutz} and then improved by Iyer and Will \cite{Iyer_1987}. For any barrier type potential whose extremities are fixed (which is our case, see FIG. \ref{fig2}), this method can be applied and yields an analytic formula that approximates the QNM frequencies.

The third order WKB approximation is given by~\cite{Iyer_1987}:
\begin{equation}\label{wkb}
	\begin{aligned}
		\omega^2_{l,r_0,n}\approx V_0&+\sqrt{-2V''_0}\Lambda\\&-i\left(n+\frac{1}{2}\right)\sqrt{-2V''_0}(1+\Omega),
	\end{aligned}
\end{equation}
with
\begin{subequations}
	\begin{equation}
		\begin{aligned}
			\Lambda=\frac{1}{\sqrt{-2V''_0}}&\Bigg[\frac{1}{8}\left(\frac{V_0^{(4)}}{V''_0}\right)\left(\frac{1}{4}+\kappa^2\right)\\&-\frac{1}{288}\left(\frac{V_0'''}{V''_0}\right)^2\left(7+60\kappa^2\right)\Bigg],
		\end{aligned}
	\end{equation}
\begin{equation}
	\begin{aligned}
		\Omega=\frac{1}{(-2V''_0)}\Bigg[&\frac{5}{6912}\left(\frac{V_0'''}{V''_0}\right)^4\left(77+188\kappa^2\right)\\
		&-\frac{1}{384}\left(\frac{V_0'''^2V_0^{(4)}}{V''^3_0}\right)\left(51+100\kappa^2\right)\\&
		+\frac{1}{2304}\left(\frac{V_0^{(4)}}{V''_0}\right)^2\left(67+68\kappa^2\right)\\
		&+\frac{1}{288}\left(\frac{V_0'''V_0^{(5)}}{V''^2_0}\right)\left(19+28\kappa^2\right)\\&
		-\frac{1}{288}\left(\frac{V_0^{(6)}}{V_0''}\right)\left(5+4\kappa^2\right)\Bigg],
	\end{aligned}
\end{equation}
\end{subequations}
where $V_0$ is the maximum value of $V_{l,r_0}$, $\kappa=n+1/2$, the primes corresponds to first, second and third order derivatives, while the superscript in round brackets~$(i)$ denotes derivative of fourth and higher orders with respect to the tortoise coordinate. In this work, the WKB approximation will be used mainly as a consistency check for the continued fraction method.

\subsection{Continued fraction method}
\label{Subsec. IV.B}

One of the most accurate methods to calculate QNMs was implemented in BH physics by Leaver \cite{Leaver_1985} and it is called the continued fraction method. This method is based on finding an analytical solution of the radial equation as a power series satisfying the QNM boundary conditions. Thus, the QNM spectrum is determined by those values of the frequencies which make the series convergent on the entire domain. 

\begin{figure}[h!]
	\centering
	\includegraphics[width=\columnwidth]{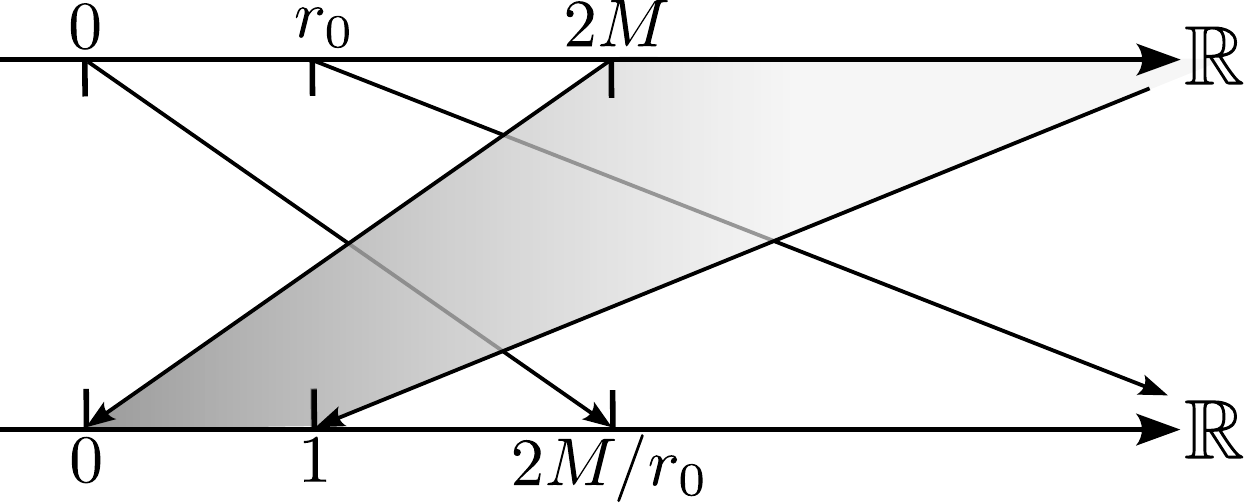}
	\caption{Schematic representation of the map defined by Eq. (\ref{map}). We assign the values taken by singularities $(0,r_0,2M,\infty)\mapsto(2M/r_0,\infty,0,1)$ and show how the domain $(2M,\infty)$ is mapped to $(0,1)$.} 
	\label{fig3}
\end{figure}

In order to apply the power series method, we first need to investigate the regular/irregular singular points of the ordinary differential equation~\eqref{radial} (see Ref.~\cite{Slavyanov} for further details). The singularities of Eq. (\ref{radial}) are $\{0,r_0,2M,\infty\}$, where the singularity at infinity is irregular and all the others are regular. The power series solutions around some singularity has a convergence radius which cannot be greater than the distance to the next neighboring singular point. Since the domain of the QNM eigenvalue problem is $(2M,\infty)$, we cannot find a well defined solution in the entire domain using a power series of $r$. Therefore, we consider the map
\begin{equation}\label{map}
	r\mapsto \frac{r-2M}{r-r_0}.
\end{equation}
Let $(0,r_0,2M,\infty)$ be the ordered 4-tuple formed by the singularities of Eq. (\ref{radial}). This 4-tuple, according to Eq. (\ref{map}), is mapped to $(2M/r_0,\infty,0,1)$ (see FIG. \ref{fig3}). Moreover, the domain $(2M,\infty)$ is compactified into $(0,1)$. The singular point $2M/r_0$ is always greater than 1, since $0<r_0<2M$. Hence, in this new coordinate defined by Eq. (\ref{map}), we can find a well defined analytical solution of Eq. (\ref{radial}) in the domain $(0,1)$, which correspond to the entire domain of interest. 


We, therefore, may consider the solution of Eq. (\ref{radial}) to be
\begin{equation}\label{solution}
	\psi_{\omega l}= r(r-r_0)^{2i \omega M +\frac{i  \omega r_0}{2}-1 }e^{i \omega  r} \sum_{n=0}^{\infty}a_n \left(\frac{r-2M}{r-r_0}\right)^{\zeta+n},
\end{equation}
where
\begin{equation}
	\zeta=\frac{- 2i \omega M }{\sqrt{1-\frac{r_0}{2M}}}
\end{equation}
is the characteristic exponent obtained from the indicial equation \cite{Slavyanov} corresponding to the ingoing solution at the horizon. The functions multiplying the summation are chosen to satisfy the boundary conditions at infinity, as well as to simplify the recurrence relation.

The sequence $(a_n)_{n\in\mathbb{N}}$ is determined by a four-term recurrence relation defined by
\begin{subequations}
	\begin{equation}
		\alpha_0 a_1+\beta_0 a_0=0
	\end{equation}    
	\begin{equation}
		\alpha_1 a_2+\beta_1 a_1+\gamma_1 a_0=0
	\end{equation}
	\begin{equation}
		\alpha_n a_{n+1}+\beta_n a_n+\gamma_n a_{n-1}+\delta_n a_{n-2}=0,\ \ \ \ n=2,3,...,
	\end{equation}    
\end{subequations}
where the recurrence coefficients are given by
\begin{widetext}
	\begin{subequations}
		\begin{equation}
			\alpha_n=-32 i \sqrt{2} M^{5/2} (n+1) \omega(2M-r_0)+8 M (n+1)^2(2M-r_0)^{3/2},
		\end{equation}    
		\begin{equation}
			\begin{aligned}
				\beta_n=&\ 64 M^4 \omega ^2(2M-r_0)^{1/2}+8 \sqrt{2} M^{5/2} \omega  (12 M \omega +12 i n+5 i)(2M-r_0)+\\&4 M \left(-2 l (l+1)+24 M^2 \omega ^2+6 i M (2 n+1) \omega -n (6 n+5)-2\right)(2M-r_0)^{3/2}+4 \sqrt{2} M^{3/2} \omega  (4 M \omega -4 i n-i)(2M-r_0)^2\\&+2 (2 n+1) (n+2 i M \omega )(2M-r_0)^{5/2},
			\end{aligned}
		\end{equation}
		\begin{equation}
			\begin{aligned}
				\gamma_n=&-128 M^4 \omega ^2(2M-r_0)^{1/2}-16 \sqrt{2} M^{5/2} \omega  (12 M \omega +6 i n-i)(2M-r_0)\\
				&+4 M \left(2 l (l+1)+n (-2-24 i M \omega )+M \omega  (-34 M \omega +3 i)+6 n^2+1\right)(2M-r_0)^{3/2}+8 \sqrt{2} M^{3/2} \omega  (6 M \omega +4 i n-i)(2M-r_0)^2\\
				&+\left(-4 l (l+1)+72 M^2 \omega ^2+4 i M (6 n-1) \omega -8 n^2+4 n-2\right)(2M-r_0)^{5/2}+8 \sqrt{2} M^{3/2} \omega ^2(2M-r_0)^3\\& +\omega  (-2 M \omega +4 i n-i)(2M-r_0)^{7/2},
			\end{aligned}
		\end{equation}    
		\begin{equation}
			\begin{aligned}
				&\delta_n=64 M^4 \omega ^2(2M-r_0)^{1/2}+8 \sqrt{2} M^{5/2} \omega  (12 M \omega +4 i n-3 i)(2M-r_0)\\
				&+4 M \left(3 n (1+4 i M \omega )+M \omega  (10 M \omega -9 i)-2 n^2-1\right)(2M-r_0)^{3/2}+\left(-4 \sqrt{2} M^{3/2} \omega  (16 M \omega +4 i n-3 i)\right)(2M-r_0)^2\\
				&+\left(n (-6-32 i M \omega )+12 M \omega  (-5 M \omega +2 i)+4 n^2+2\right)(2M-r_0)^{5/2}+8 \sqrt{2} M^{3/2} \omega ^2(2M-r_0)^3\\
				&+\omega  (14 M \omega +4 i n-3 i)(2M-r_0)^{7/2} -\omega ^2(2M-r_0)^{9/2}.
			\end{aligned}
		\end{equation}
	\end{subequations}
\end{widetext}

This recurrence relation is in agreement with Leaver's hypothesis, which says that a radial equation with a confluent singularity and three regular singularities generates a solution whose expansion coefficients obey a 4-term recurrence relation \cite{Leaver_1990}.

To calculate the QNM from a 4-term recurrence relation we first have to apply the Gaussian elimination scheme, defined by
\begin{equation}
	\tilde{\alpha}_n \equiv \alpha_n,\quad \tilde{\beta}_n\equiv \beta_n,\quad \tilde{\gamma}_n\equiv \gamma_n,\quad \text{for}\ n=0,1\ ,
\end{equation}
and
\begin{subequations}
	\begin{align}
		&\tilde{\delta}_n\equiv 0, \quad \tilde{\alpha}_n\equiv \alpha_n,\\
		&\tilde{\beta}_n\equiv \beta_n-\frac{\tilde{\alpha}_{n-1}\delta_n}{\tilde{\gamma}_{n-1}}, \quad \tilde{\gamma}_n\equiv \gamma_n-\frac{\tilde{\beta}_{n-1}\delta_n}{\tilde{\gamma}_{n-1}}\ \ \ \text{for}\ n\geq2.
	\end{align}
\end{subequations}

\begin{figure*}
	\centering
	\includegraphics[scale=1]{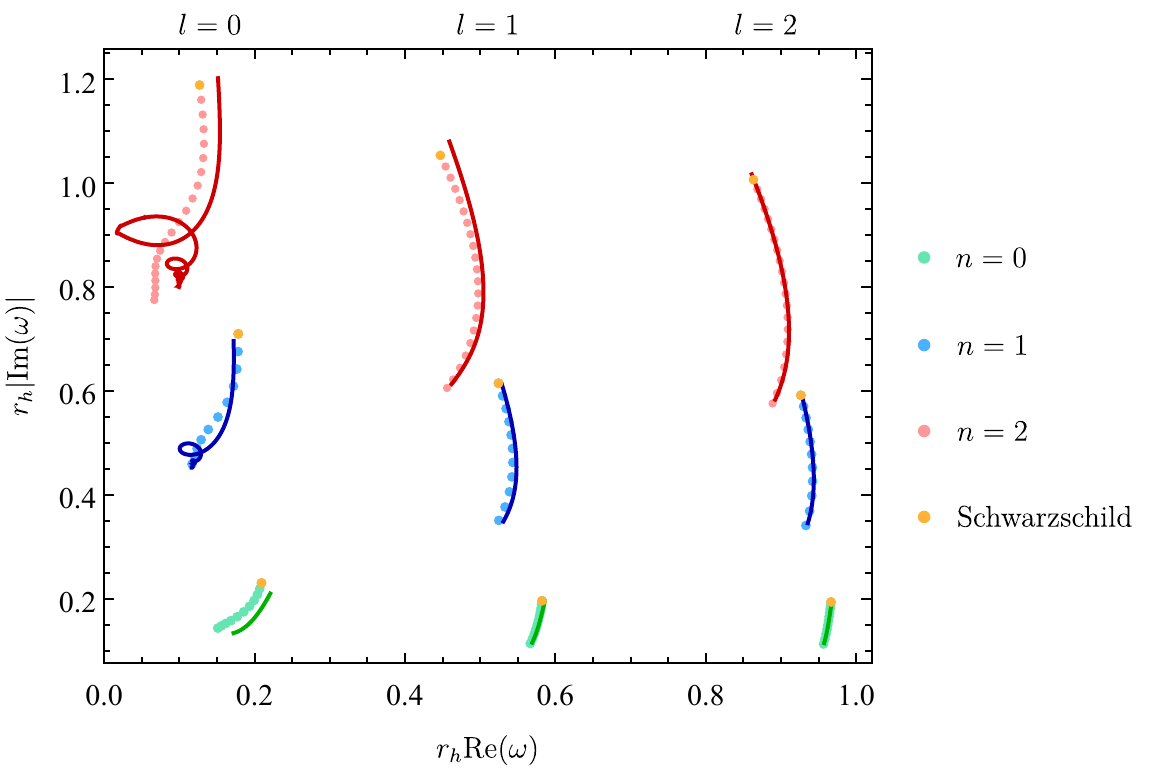}
	\includegraphics[scale=1]{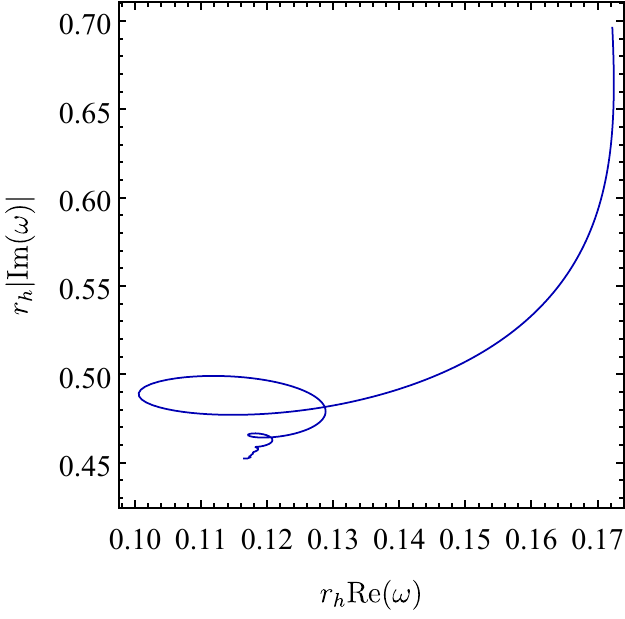}
	\includegraphics[scale=0.99]{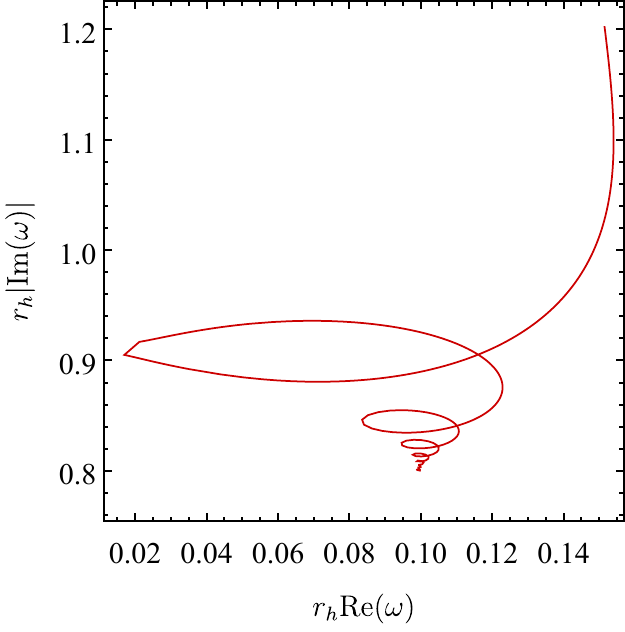}
	\caption{\textit{Top}: First three ($n=0,1,2$) QNM-frequencies of the scalar field on the quantum corrected Schwarzschild spacetime. The green plots correspond to $n=0$, blue to $n=1$ and red to $n=2$. Circles represent the WKB calculations, for several values of the LQG parameter $r_0$, beginning at $r_0/r_h=0$ (yellow circles representing the Schwarzschild case) and ending at $r_0/r_h=0.99$. The solid line shows the continued fraction calculation in the same range of parameters. \textit{Bottom}: Computation of QNM frequencies for $n=1$, $l=0$ (bottom left panel) and $n=2$, $l=0$ (bottom right panel) for $r_0/r_h$ ranging from 0 to 0.99, obtained using the continued fraction method with Nollert improvement \cite{Nollert_1993}. All the frequencies become less damped as $r_0/r_h$ increases.}
	\label{WKB}
\end{figure*}

The new recurrence coefficients now obey a 3-term recurrence relation given by
\begin{subequations}\label{3rec}
	\begin{equation}
		\tilde{\alpha}_0a_1+\tilde{\beta}_0a_0=0,
	\end{equation}
	\begin{equation}
		\tilde{\alpha}_na_{n+1}+\tilde{\beta}_na_n+\tilde{\gamma}_na_{n-1}=0,\ \ n=1,2,...\ .
	\end{equation}
\end{subequations}

The condition that the series defined in Eq. (\ref{solution}) converges uniformly is given by \cite{Leaver_1985}
\begin{equation}\label{continued}
	\begin{aligned}
		0&=\tilde{\beta}_0-\cfrac{\tilde{\alpha}_0\tilde{\gamma}_1}{\tilde{\beta}_1-\cfrac{\tilde{\alpha}_1\tilde{\gamma}_2}{\tilde{\beta}_2-\cfrac{\tilde{\alpha}_2\tilde{\gamma}_3}{\tilde{\beta}_3-...}}}, \\
		&\equiv \tilde{\beta}_0-\frac{\tilde{\alpha}_0\tilde{\gamma}_1}{\tilde{\beta}_1-}\frac{\tilde{\alpha}_1\tilde{\gamma}_2}{\tilde{\beta}_2-}\frac{\tilde{\alpha}_2\tilde{\gamma}_3}{\tilde{\beta}_3-}...
	\end{aligned}
\end{equation}
Thus, the set of frequencies that makes Eq. (\ref{continued}) true are, precisely, the QNM frequencies. 

The roots of Eq.~(\ref{continued}) can be found numerically. The most stable root of the continued fraction defined in Eq.~(\ref{continued}) is the fundamental mode. 
The $n$-th inversion of Eq. (\ref{continued}) is defined by
\begin{equation}\label{inversion}
	\tilde{\beta}_n-\frac{\tilde{\alpha}_{n-1}\tilde{\gamma}_n}{\tilde{\beta}_{n-1}-}
		...-\frac{\tilde{\alpha}_0\tilde{\gamma}_1}{\tilde{\beta}_0}=
		\frac{\tilde{\alpha}_{n}\tilde{\gamma}_{n+1}}{\tilde{\beta}_{n+1}-}\frac{\tilde{\alpha}_{n+1}\tilde{\gamma}_{n+2}}{\tilde{\beta}_{n+2}-}...,
\end{equation}
and its most stable root is the $n$-th eigenfrequency.

\subsection{Prony method}
\label{Subsec. IV.C}

We can also solve Eq. \eqref{kg} without assuming the time dependence $e^{-i \omega t}$ in Eq. \eqref{field}. This lead to the partial differential equation (PDE):

\begin{equation}\label{pde}
	\frac{\partial^2 \Psi_{ l}}{\partial r_*^2}-\frac{\partial^2 \Psi_{l}}{\partial t^2}-V_{l,r_0}[r(r_*)]\Psi_{ l}=0,
\end{equation}
where now $\Psi_l$ is a function of the variables $(t,r_{*})$. We may solve Eq. (\ref{pde}) numerically, setting a Gaussian wave package centered at $r_{*}=0$ as our initial configuration for the field. The time evolution of the solution is characterized by three stages: \text{(i):} a prompt response at early times, which is strongly determined by the chosen initial conditions of the field, \textit{(ii):} exponential decay at intermediate times, determined by the QNMs and \textit{(iii):} power-law fall-off at late times, due to backscattering of the field in tail of the potential. 

Once the Eq. (\ref{pde}) is solved for some initial data configuration, by means of estimating methods, one is able to construct an analytic approximation that fits the original solution. Here we use the Prony method to find an approximate Fourier decomposition, which allows us to calculate the fundamental mode. A detailed description of the Prony method can be found in Ref. \cite{Berti_2007}.

\section{Results}
\label{Sec. V}

 In this section we exhibit a selection of our results, obtained from the methods described in the previous sections.

 As a consistency check that the continued fraction method leads to the correct values of the eigenfrequencies, we first compare the QNMs calculated with Leaver's method and the WKB approximation.

We compute the QNMs $n=0,1,2$ for different numbers of the azimuthal number $l=0,1,2$. The results are exhibited in the top panel of FIG.~\ref{WKB}. We define a color code for each value of $n$, namely: green $\leftrightarrow n=0$, blue $\leftrightarrow n=1$ and red $\leftrightarrow n=2$. The (green, blue and red) circles represent the QNMs calculated with the WKB method, while the solid lines were obtained by the continued fraction method. Both, WKB and continued fraction calculations, were computed for $r_0/r_h$ ranging from 0 (Schwarzschild) to 0.99. The yellow circles, located at the top of each continuous line, represents the quasinormal frequencies of the Schwarzschild BH calculated using the WKB approximation.
 
FIG.~\ref{WKB} shows that, as we increase the values of $l$, the results obtained from WKB and continued fraction methods converge to the same value, which was already expected. Nonetheless, even for $l=0$, both, WKB and continued fraction methods, results are in very good agreement. 

From the continued fraction computations we also note that the curves in the complex plane, parametrized by $r_0/r_h$, for $n>0$ and $l=0$, have a spiral-like shape. 
We display the curves for $n=1$, $l=0$ and $n=2$, $l=0$ in the left and right bottom panels of FIG.~~\ref{WKB}, respectively. We note that, in order to obtain the results shown in the bottom panel of FIG.~\ref{WKB}, we applied the continued fraction method with the improvement proposed by Nollert~\cite{Nollert_1993}. The Nollert improvement is suitable to compute QNM frequencies with large imaginary part, hence it gives accurate numerical results when the LQG parameter is close to the extreme value. While the LQG parameter varies in the indicated range, the trajectory described in the complex plane moves away from the Schwarzschild QNMs and spirals towards some fixed complex value, which corresponds to the QNMs associated with the extremal case. A similar behavior was also found for the Reissner-Nordström BH \cite{Berti_2003}. We also note that that these curves are self-intersecting. The existence of self-intersecting curves in the orbits of the QNM frequencies is related to the fact that, for different values of the LQG parameter $r_0$, the BH may present the same frequency for some given $n$. 


\begin{figure*}
	\centering
	\includegraphics[scale=1]{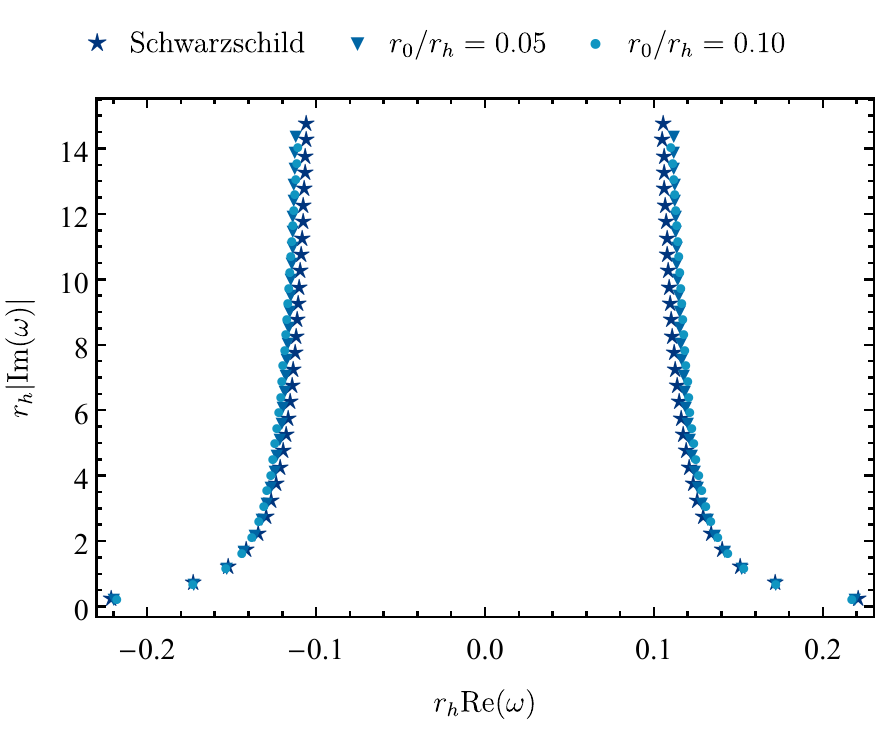}
	\includegraphics[scale=1]{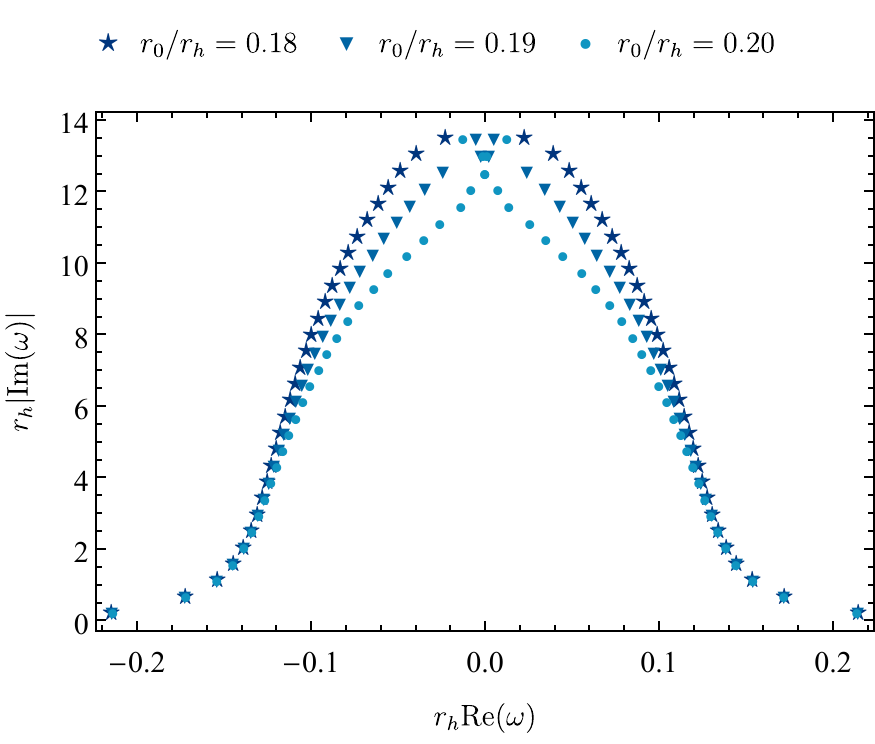}
	\includegraphics[scale=1]{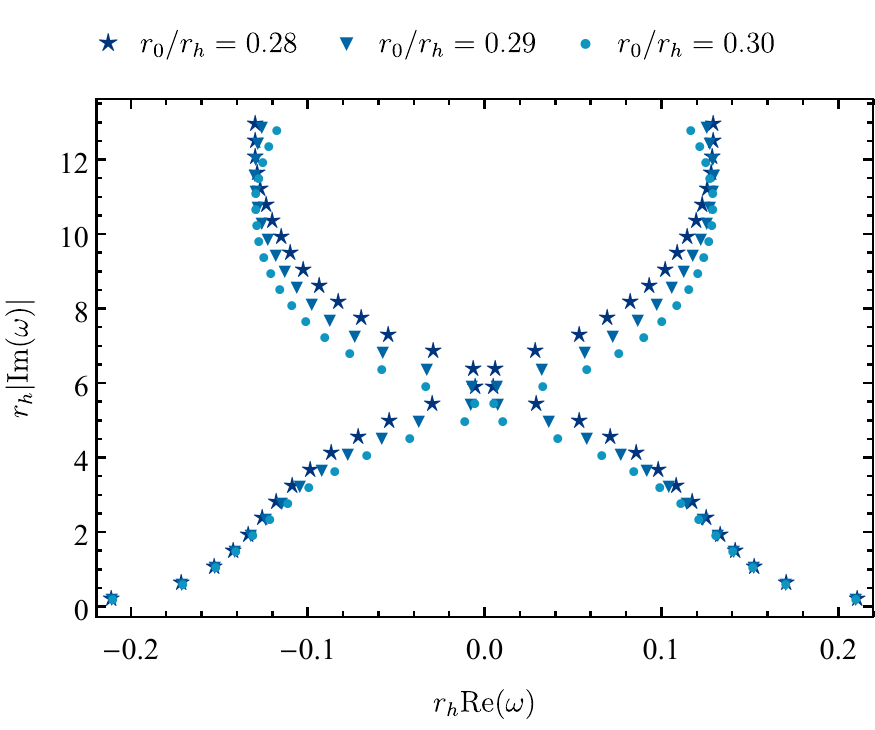}
	\includegraphics[scale=1]{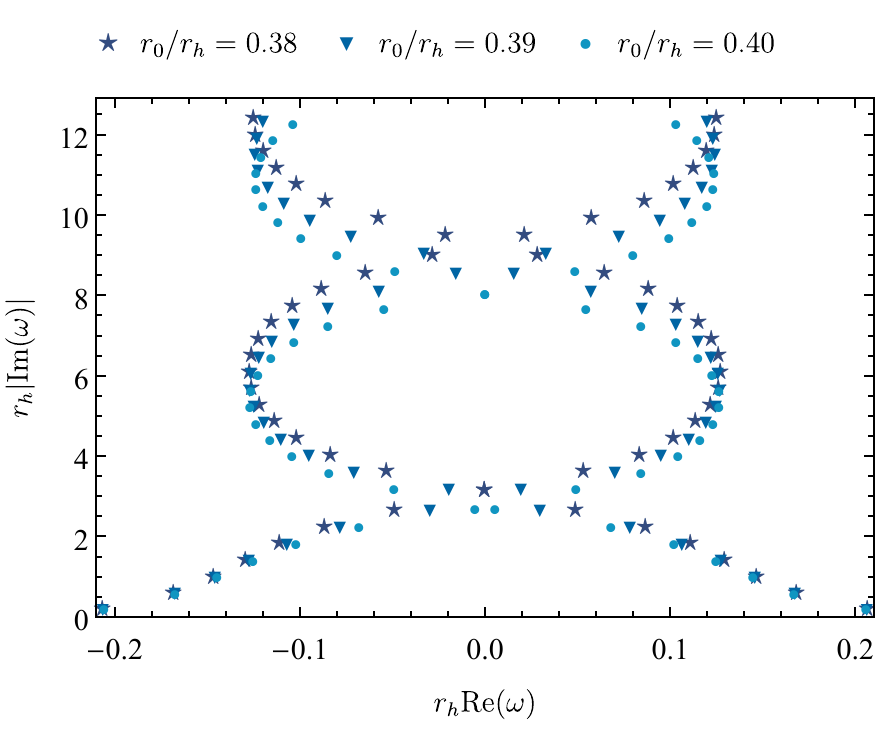}
	\includegraphics[scale=1]{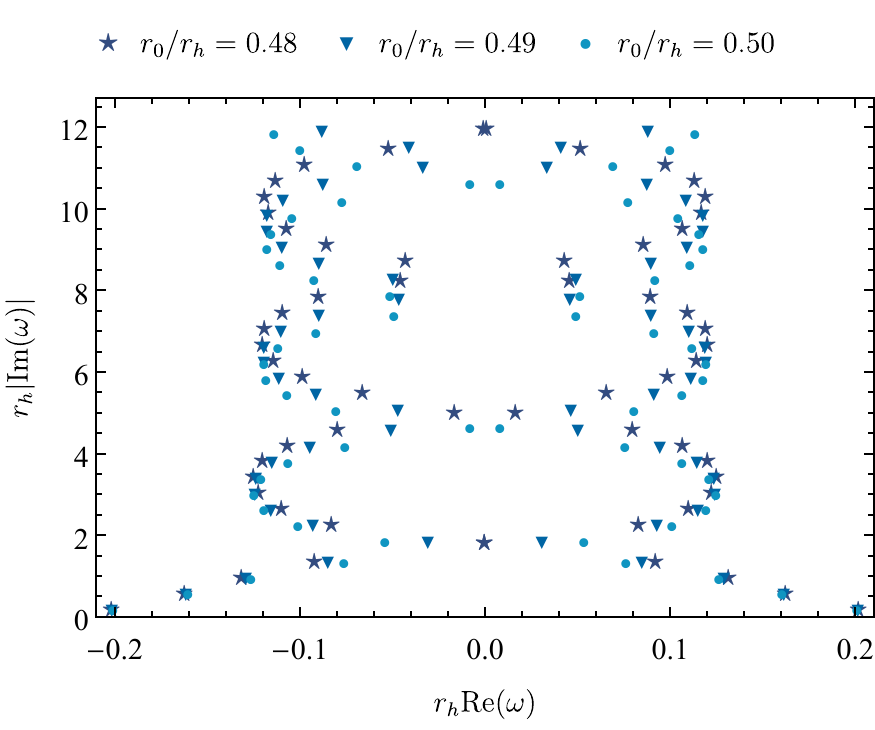}
	\caption{First 30 QNM frequencies ($n=0,1,...,29$) of the scalar field on the quantum corrected Schwarzschild spacetime for $l=0$. The top left panel shows the modes for $r_0/r_h=0,\ 0.05,\ 0.1$, whereas the remaining panels exhibit the spectrum of the quantum corrected BH for values of the LQG parameter near $r_0/r_h=0.2,\ 0.3,\ 0.4, \ 0.5$. All the spectra were calculated with the continued fraction method.}
	\label{specl0}
\end{figure*}



 \begin{figure*}
	\centering
	\includegraphics[scale=1]{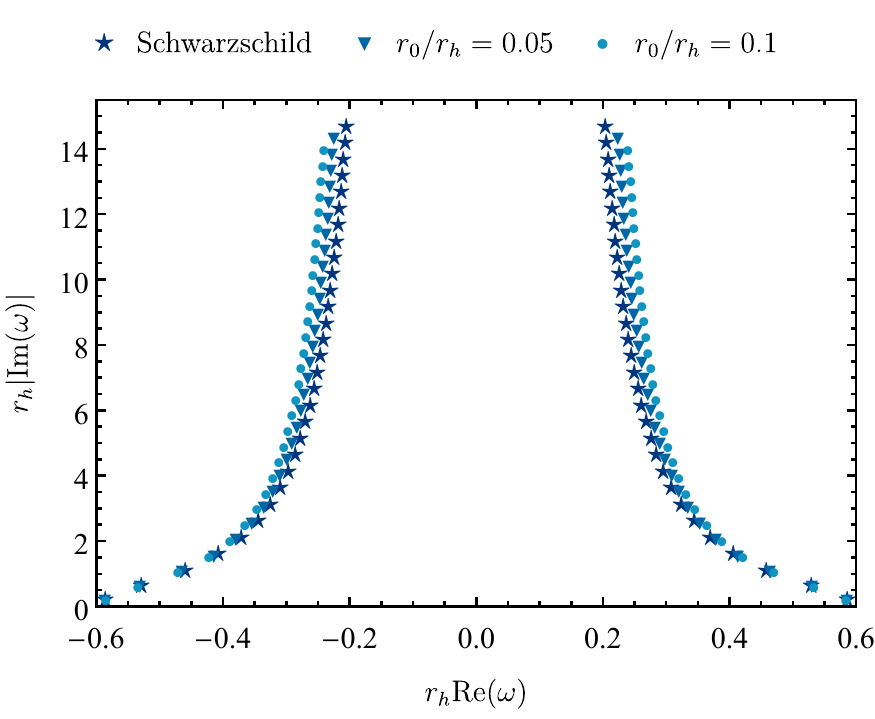}
	\includegraphics[scale=1]{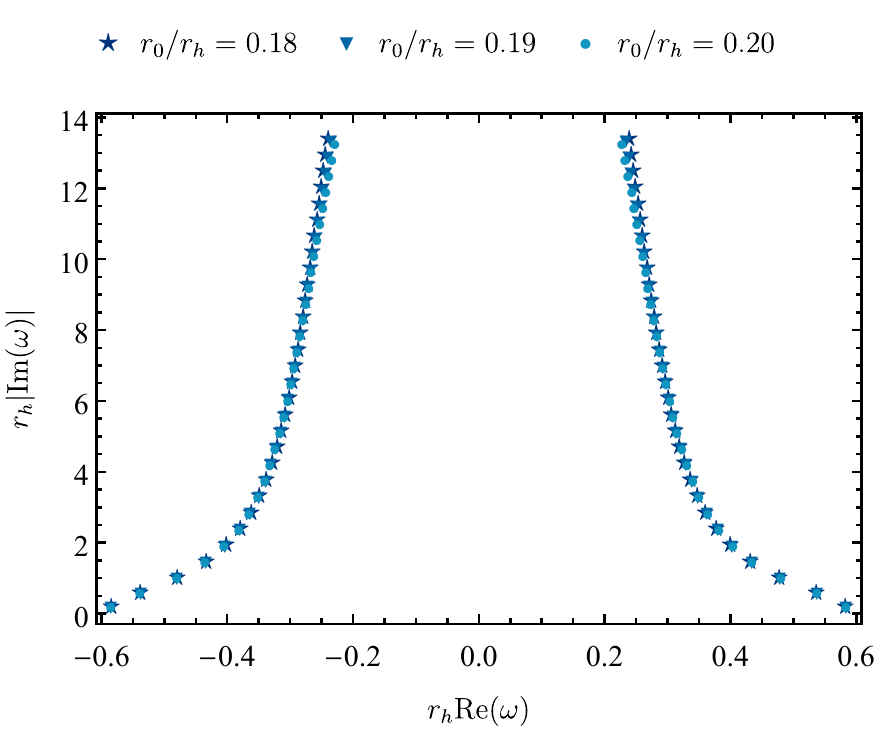}
	\includegraphics[scale=1]{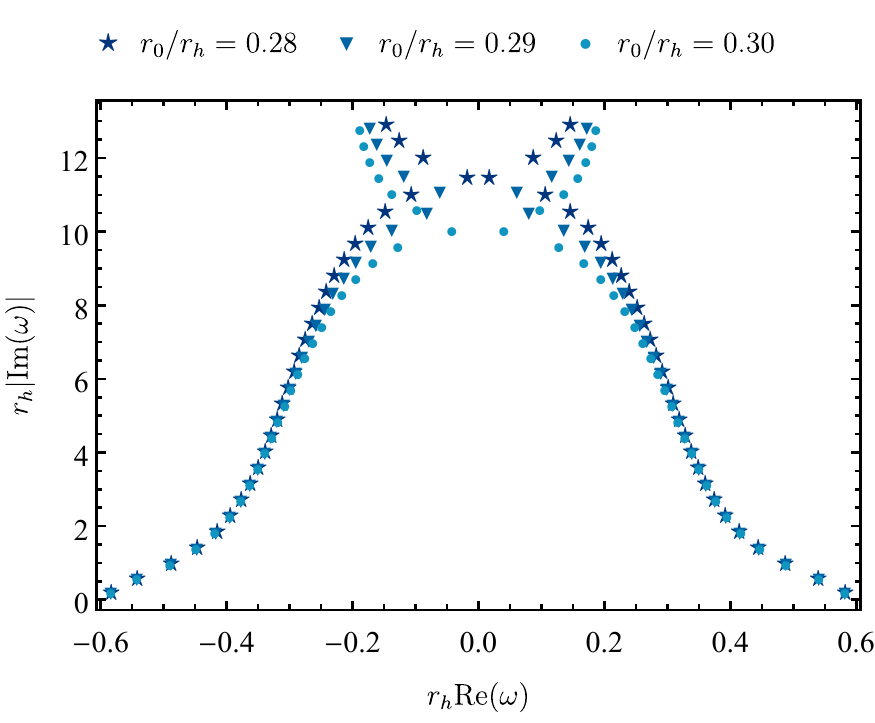}
	\includegraphics[scale=1]{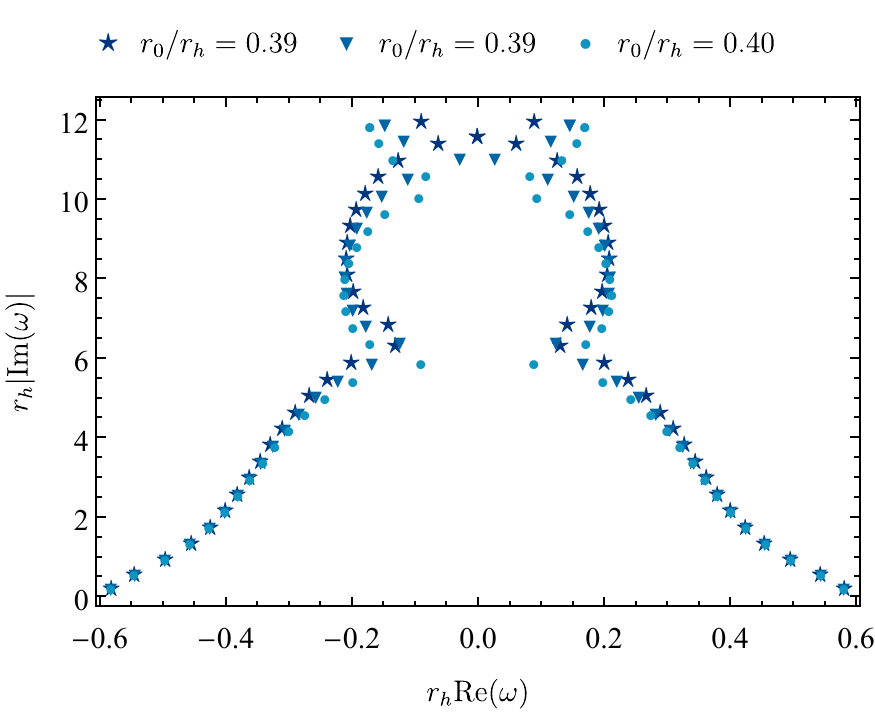}
	\includegraphics[scale=1]{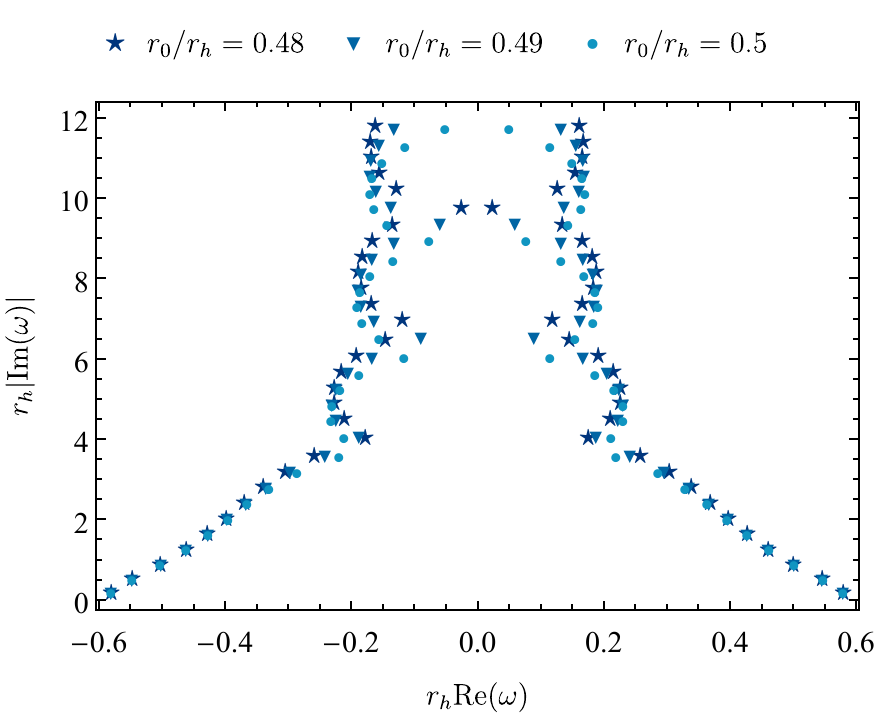}
	\caption{First 30 QNM frequencies ($n=0,1,...,29$) of the scalar field on the quantum corrected Schwarzschild spacetime for $l=1$. We display in the top left panel the modes $r_0/r_h=0,\ 0.05,\ 0.1$, whereas in the remaining panels we show the spectrum for higher values of the LQG parameter, namely near $r_0/r_h=0.2,\ 0.3,\ 0.4, \ 0.5$. All spectra were calculated with the continued fraction method.}
	\label{specl1}
\end{figure*}

\subsection{$l=0$ Modes}
\label{SecVA}

We can fix $l=0$ and compute the first 30 modes for several values of $r_0/r_h$. The results, calculated with the continued fraction method, are displayed in FIG. \ref{specl0}.  The small deviation from Schwarzschild regime ($r_0/r_h=0,\ 0.05,\ 0.1$) is displayed in the top left panel of FIG. \ref{specl0}, where we obtain the famous Schwarzschild's scalar spectrum, formed by two non-intersecting  branches of QNM frequencies, with slight disturbances. There is a decrease in the damping, in accordance with FIG.~\ref{WKB}. 

Nonetheless, the QNM frequencies for higher values of $r_0/r_h$ are completely different from the Schwarzschild case. In the remaining panels of FIG. \ref{specl0} we exhibit the spectrum near $r_0/r_h=0.2,\ 0.3,\ 0.4, \ 0.5$. As the LQG parameter varies, the real part of the frequencies oscillates. We obtained frequencies with $\text{Re}(\omega)=0$, e.g. the mode $n=4$ for $r_0/r_h=0.48$. However, the continued fraction method does not converge for $\text{Re}(\omega)\rightarrow0$, hence the existence of purely damped modes cannot be indeed stated.

We remark that the existence of frequencies with real part almost equal to zero can be found in Schwarzschild's QNM spectrum for the gravitational field. The algebraically special frequency $2M\omega\approx-i(l-1)(l+1)(l+2)/6$ is almost a pure imaginary number \cite{Chandrasekar_1984}. In the Schwarzschild case the algebraically special frequency does not exist for fields other than the gravitational field. Thus, qualitatively, in the weakly damped regime, the spectrum of the quantum corrected Schwarzschild BH for the scalar field resembles the spectrum of the Schwarzschild BH for the gravitational field.

\begin{table}[h]
	\begin{tabular}{ |c||c|c|  }
		
		\hline
		\multicolumn{3}{|c|}{$l=0$ (Leaver)} \\
		\hline
		$n$& Schwarzschild& $r_0/r_h=  0.3$ \\
		\hline
		0   & $0.2209-0.2097i$& $0.2099-0.1828i$    \\
		\hline
		1&   $0.1722-0.6961i$& $0.1705-0.5996 i$  \\
		\hline
		2 &$0.1514-1.2021 i$& $0.1518-1.0329 i$ \\
		\hline
		3    &$0.1408-1.7073 i$& $0.1403-1.4657 i$ \\
		\hline
		4&   $0.1341-2.2112 i$& $0.1307-1.8974 i$ \\
		\hline
	\end{tabular}
	\caption{First five overtones of scalar perturbations, expressed in $r_h^{-1}$ units, calculated by the continued fraction method for $l=0$. We consider the Schwarzschild BH, as well as the holonomy corrected Schwarzschild BH with $r_0/r_h=  0.3$.}
	\label{table1}
\end{table}

\begin{figure*}
 	\centering
 	\includegraphics[scale=1]{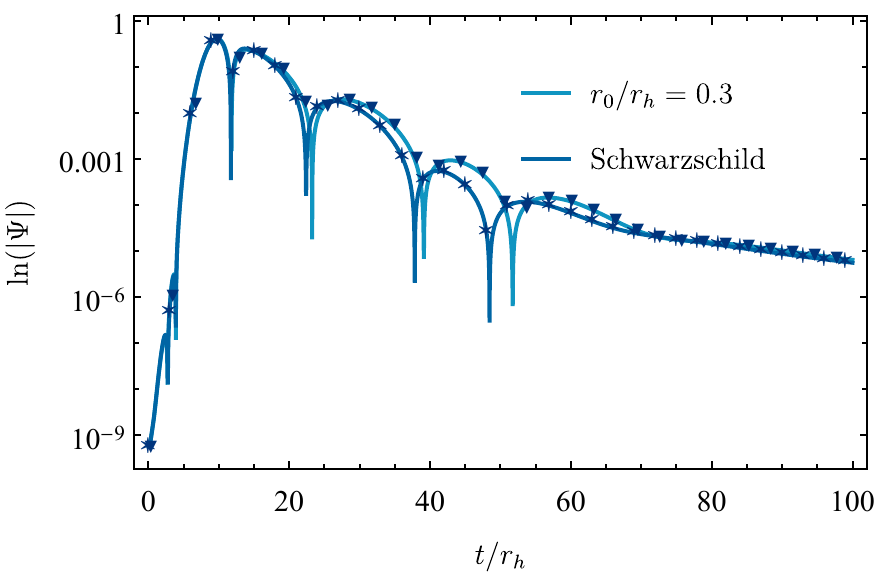}
 	\includegraphics[scale=1]{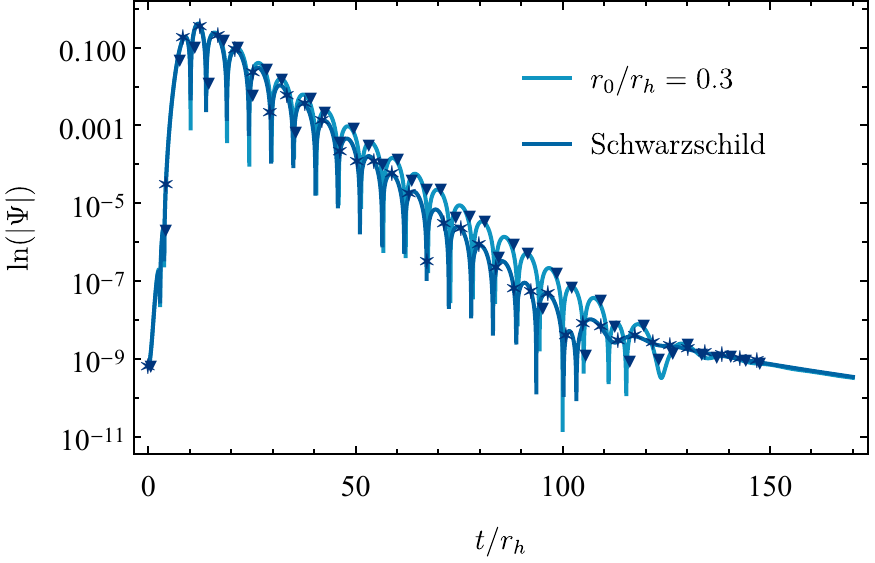}
 	\caption{Plots of $\ln|\Psi_l(t,r_{*}=10)|$ for $l=0$ (left panel) and $l=1$ (right panel). In both plots, we include the Schwarzschild case (darker blue), the quantum corrected Schwarzschild case with $r_0/r_h=0.3$ (lighter blue) and also their associated fittings obtained from the Prony method (stars and inverted triangles, respectively).}
 	\label{tail}
 \end{figure*}

 \subsection{$l=1$ Modes}
 
 We may now fix $l=1$ and compute the first 30 modes for the same values of the LQG parameter of Sec.~\ref{SecVA}. The QNMs were calculated using the continued fraction method and are displayed in FIG. \ref{specl1}. Once more, small holonomy corrections $r_0/r_h=0,\ 0.05,\ 0.1$  lead to small disturbances in the Schwarzschild spectrum, leading to an overall decrease in the imaginary part of the QNM frequencies.
 
The QNMs for higher values of the LQG parameter, namely, near $r_0/r_h=0.2,\ 0.3,\ 0.4, \ 0.5$, are displayed in the remaining panels of FIG.~\ref{specl1}. As we increase the value of $r_0/r_h$, again the oscillatory pattern appears in the spectrum of the quantum corrected BH. 
The first five overtones are also exhibited in Table \ref{table2}.

 \begin{table}[h]
 	\begin{tabular}{ |c||c|c|  }
 		
 		\hline
 		\multicolumn{3}{|c|}{$l=1$ (Leaver)} \\
 		\hline
 		$n$& Schwarzschild& $r_0/r_h=  0.3$ \\
 		\hline
 		0   & $0.5858-0.1953i$& $0.5824-0.1738i$    \\
 		\hline
 		1&   $0.5288-0.6125i$& $0.5416-0.5395 i$  \\
 		\hline
 		2 &$0.4590-1.0802 i$& $0.4892-0.9406 i$ \\
 		\hline
 		3    &$0.4065-1.5766 i$& $0.4473-1.3640 i$ \\
 		\hline
 		4&   $0.3702-2.0815 i$& $0.4172-1.7952 i$ \\
 		\hline
 	\end{tabular}
 	\caption{First five overtones of scalar perturbations, expressed in $r_h^{-1}$ units, calculated by the continued fraction method for $l=1$. We considered the Schwarzschild BH, as well as the holonomy corrected Schwarzschild BH with $r_0/r_h=  0.3$.}
 	\label{table2}
 \end{table}

  The pattern exhibited in both FIGs.~\ref{specl0} and ~\ref{specl1} may go on forever as $n$ increases. However such analysis requires an asymptotic study of QNMs that is beyond the scope of this paper. If this assertion is true, then the limit $\lim_{n\rightarrow\infty}\Re \omega$ does not exist, what would differ from the Schwarzschild's case, which is known to be $2M\lim_{n\rightarrow\infty}\Re \omega= \ln3/4\pi$ \cite{Motl_2003}. 

\subsection{Time domain profile}

We may solve numerically the time-dependent wave equation given by Eq.~(\ref{pde}), and for that, we need to specify an initial condition. We consider the initial data as a Gaussian, according to
\begin{equation}\label{gaussian}
	\Psi_l(0,r_{*})=e^{-r_{*}^2/4};\ \partial_t\Psi_l(t,r_{*})|_{t=0}=0.
\end{equation}
The chosen initial data does not play a significant role in the time profile of the wave function from intermediate times onwards. After a transient initial stage (highly dependent on initial conditions), the time profile is dominated by the QNMs and then by the late time tail decay.

In FIG. \ref{tail} we exhibit the logarithmic plot for the  absolute value of the solution as a function of time. The tortoise coordinate is fixed at $r_{*}/r_h=10$. We consider the cases $l=0$ (left panel) and $l=1$ (right panel). The logarithmic wave-form is calculated for the cases of Schwarzschild (darker blue) and holonomy corrected Schwarzschild with $r_0/r_h=0.3$ (lighter blue). We also include the respective Prony's fittings (stars for Schwarzschild and inverted triangles for quantum corrected Schwarzschild).

The three phases described in the Sec. \ref{Subsec. IV.C} are clearly distinguished in FIG. \ref{tail}. We highlight that the power-law tail developed at late times seems to be independent of the loop quantum correction.

 \begin{table}[h!]
	\begin{tabular}{ |c||c|c|  }
		
		\hline
		\multicolumn{3}{|c|}{$l=0$ (Prony)} \\
		\hline
		$n$& Schwarzschild& $r_0/r_h=  0.3$ \\
		\hline
		0   & $0.2209 -0.2098 i$& $0.2098 -0.1828 i$    \\
		\hline
	\end{tabular}
	\begin{tabular}{ |c||c|c|  }
	
	\hline
	\multicolumn{3}{|c|}{$l=1$ (Prony)} \\
	\hline
	$n$& Schwarzschild& $r_0/r_h=  0.3$ \\
	\hline
	0   & $0.5858 -0.1953 i$& $0.5823 -0.1737 i$    \\
	\hline
\end{tabular}
	\caption{Fundamental frequency of scalar perturbations, expressed in $(2M)^{-1}$ units, calculated by the Prony method for $l=0,1$. We considered the Schwarzschild BH, as well as the holonomy corrected Schwarzschild BH with $r_0/r_h=  0.3$.}
	\label{table3}
\end{table}

The fundamental modes, obtained from the Prony method, for $l=0,1$ are showed in TABLE \ref{table3}. This results can be compared with those of TABLES \ref{table1} and \ref{table2}. As we can see, both results are in excellent agreement.

\section{DISCUSSION AND CONCLUSIONS}
\label{Sec. VI} 
Computing QNMs of BHs is a long-standing task in BH physics.
The first calculation of QNMs as a boundary problem was carried out in a paper by  Chandrasekhar and Detweiler \cite{Chandrasekhar_1975}. Later on, Leaver developed a simple, but very powerful approach to this problem \cite{Leaver_1985,Leaver_1990}. Since then, the calculation of QNMs of different types of astrophysical objects has been carried out. 
Most of the main BHs spacetimes already have their QNMs cataloged in several tables with great precision.

Simultaneously with the progress of the BH perturbation theory, the search for a quantum theory of gravity was strongly active. Among several possibilities, the theory of LQG has had many interesting results, namely, the construction of singular-free cosmological and BH solutions \cite{Bojowald_2001,Gambini_2013} and the derivation of the Hawking-Bekenstein entropy \cite{Ashtekar_1998}. These results might be the smoking guns to a complete and consistent theory of quantum gravity. 
Nevertheless, there is still a lot of work to be done until we can interpret all LQG results properly. Thus, many effective models have been studied, aiming to obtain effects that one would expect to observe in the complete LQG theory.

We investigated the scalar QNMs of a quantum corrected Schwarzschild BH. We used standard methods of BH perturbation theory, namely, the third order WKB approximation, the continued fraction method (also named as Leaver's method) and the Prony method. In order to perform a consistency check, we compared the numerical results, computed through the three different methods, and obtained an excellent agreement, in the regime of applicability of each method. 

We computed the QNMs for different values of the multipole number $l$ and the overtones $n$. In particular, we obtained the first 30 overtones for the fundamental mode $l=0$ and the first 30 overtones for the mode $l=1$, using the Leaver's method. Our numerical results show that, for a fixed $l$ and $n$, the quantum corrected Schwarzschild BH perturbations become less damped as we increase the LQG parameter $r_0$. Moreover, for $l=0$ and $n>0$, the QNMs frequencies curves in the complex plane are self intersecting, meaning that two different quantum corrected Schwarzschild BH configurations may have the same QNMs.

Furthermore, we obtained that for middle-to-high values of $r_0/r_h$, the scalar QNMs of the quantum corrected Schwarzschild BH may have vanishing real part, i.e. it admits purely decaying modes. We remark that purely decaying modes in a classical Schwarzschild BH exist solely for gravitational perturbations~\cite{Chandrasekar_1984,Leaver_1985}.

The recent detection of GWs has deepened our understanding of the classical nature of gravity. It is possible that future generations of GW detectors, such as the LISA detector, can probe the quantum nature of the gravitational field. In this work, we obtained that the QNM oscillations of a quantum-corrected BH can be very different from the Schwarzschild one. Our results indicate that the story about the quantum nature of gravity can be heard from the sounds played by a BH.

\bigskip

{\bf Note added:}  As this paper was being completed, we became aware of Ref.~\cite{MN1}, which also covers, albeit using different techniques, the scalar QNMs of the spacetime~(\ref{metric}), but in a different range of the parameter space and a more restrictive set of overtone numbers. In both cases, our analysis required specific techniques, such as the Leaver's method with Nollert improvement.

\begin{acknowledgments}

We are grateful to Funda\c{c}\~ao Amaz\^onia de Amparo a Estudos e Pesquisas (FAPESPA), Conselho Nacional de Desenvolvimento Cient\'ifico e Tecnol\'ogico (CNPq) and Coordena\c{c}\~ao de Aperfei\c{c}oamento de Pessoal de N\'ivel Superior (CAPES) -- Finance Code 001, from Brazil, for partial financial support. 
ZM, HLJ and LC thank the University of Aveiro, in Portugal, for the kind hospitality during the completion of this work.
This work is supported  by the  Center for Research and Development in Mathematics and Applications (CIDMA)
through the Portuguese Foundation for Science and Technology (FCT -- Fundaç\~ao para a Ci\^encia e a 
Tecnologia), references  UIDB/04106/2020 and UIDP/04106/2020.  
The authors acknowledge support  from the projects CERN/FIS-PAR/0027/2019, PTDC/FIS-AST/3041/2020, 
CERN/FIS-PAR/0024/2021 and 2022.04560.PTDC.  
This work has further been supported by  the  European  Union's  Horizon  2020  research  and  innovation
(RISE) programme H2020-MSCA-RISE-2017 Grant No.~FunFiCO-777740 and by the European Horizon Europe staff
exchange (SE) programme HORIZON-MSCA-2021-SE-01 Grant No.~NewFunFiCO-101086251.
\end{acknowledgments}


\begin{thebibliography}{99}

	
	\bibitem{Ligo_a}{The LIGO Scientific Collaboration, and The Virgo Collaboration, Observation
		of gravitational waves from a binary black, Phys. Rev. Lett. \textbf{116}, 061102 (2016).}
	
	\bibitem{Ligo_b}{The LIGO Scientific Collaboration, and The Virgo Collaboration, GW151226: Observation of gravitational waves from a 22-solar-mass binary black
		hole coalescence, Phys. Rev. Lett. \textbf{116}, 241103 (2016).}
	
	\bibitem{Kerr_1963}{R. P. Kerr, Gravitational field of a spinning mass as an example of algebraically special metrics, Phys. Rev. Lett. \textbf{11}, 237 (1963).}
	
	\bibitem{Pretorious_2005}{F. Pretorius, Evolution of Binary Black-Hole Spacetimes, Phys. Rev. Lett. \textbf{95}, 121101 (2005).}
	
	\bibitem{Campanelli_2006}{M. Campanelli, C. O. Lousto, P. Marronetti, and Y. Zlochower, Accurate evolutions of orbiting black-hole binaries without excision , Phys. Rev. Lett. \textbf{96,}, 111101 (2006).}
	
	\bibitem{Baker_2006}{J. G. Baker, J. Centrella, D-I. Choi, M. Koppitz, and J. van Meter, Phys. Rev. Lett. \textbf{96}, 111102 (2006).}
	
	\bibitem{Echeverria_1988}{F. Echeverria, Gravitational-wave measurements of the mass and angular momentum of a black hole, Phys. Rev. D \textbf{40}, 3194 (1989).}
	
	\bibitem{Finn_1992}{L. S. Finn, Detection, measurement, and gravitational radiation, Phys. Rev. D \textbf{46}, 5236 (1992).}
	
	\bibitem{Berti_2006}{E. Berti, V. Cardoso, and C. M. Will, Gravitational-wave spectroscopy of massive black holes with the space interferometer LISA, Phys. Rev. D \textbf{73}, 064030 (2006).}
	
	\bibitem{Berti_2009}{E. Berti, V. Cardoso, and A. O Starinets, Quasinormal modes of black holes and black branes, Class. Quantum Grav. \textbf{26}, 163001 (2009).}
	
	\bibitem{Kokkotas_1999}{K. D. Kokkotas, and B.G Schmidt, Quasi-normal modes of stars and black holes, Living Rev. Relativ. \textbf{2}, 2 (1999).}
	
	\bibitem{Regge_1957}{T. Regge, and J. A. Wheeler, and B.G Schmidt, Stability of a Schwarzschild singularity, Phys. Rev. \textbf{108}, 1063 (1957).}
	
	\bibitem{Zerilli_1970a}{F. J. Zerilli, Effective potential for even-parity Regge-Wheeler gravitational perturbation equations, Phys. Rev. Lett.  \textbf{24}, 737 (1970).}
	
	\bibitem{Zerilli_1970b}{F. J. Zerilli, Gravitational field of a particle falling in a Schwarzschild geometry analyzed in tensor harmonics, Phys. Rev. D \textbf{2}, 2141 (1970).}
	 
	\bibitem{Chandrasekhar_1975}{S. Chandrasekhar, and S. Detweiler, The quasi-normal modes of the Schwarzschild black hole,  Proc. R. Soc. Lond. A \textbf{344}, 411–452 (1975).}
	
	\bibitem{Lemos}{N. A. Lemos, \textit{Convite à física matemática} (Editoria livraria da física, São Paulo, 2013).}
	
	\bibitem{Konishi}{K. Konishi, and G. Paffuti \textit{Quantum mechanics: A new introduction} (Oxford University Press, New York, 2009).}
	
	\bibitem{dinverno}{R. D'Inverno, \textit{Introducing Einstein's Relativity} (Oxford University Press, Oxford, 1899).}
		
	\bibitem{Will}{C. M. Will, The confrontation between general relativity and experiment, Living Rev. Relativ. \textbf{17}, 4 (2014).}
	
	\bibitem{M87}{The Event Horizon Telescope Collaboration, First M87 Event Horizon Telescope Results. I. The Shadow of the Supermassive Black Hole, ApJL. \textbf{875}, L1 (2019).}
	
	\bibitem{Sgr}{The Event Horizon Telescope Collaboration, First Sagittarius A* Event Horizon Telescope Results. I. The Shadow of the Supermassive Black Hole in the Center of the Milky Way, ApJL. \textbf{930}, L12 (2022).}
	
	\bibitem{Hawking_1970}{S. W. Hawking, and R. Penrose, The singularities of gravitational collapse and cosmology, Proc. R Soc. Lond. Ser A. \textbf{314}, 529-548 (1970).}
	
	\bibitem{Penrose_1965}{Gravitational collapse and space-time singularities, Phys. Rev. Lett. \textbf{14}, 57 (1965).}
	
	\bibitem{Hawking_1975}{S. W. Hawking, Particle creation by black holes, Comm. Math. Phys. \textbf{43}, 199 (1975).}
	
	\bibitem{DeWitt_1967}{B. S. DeWitt, Quantum theory of gravity. I. The canonical theory, . \textbf{160}, 1113 (1967).}
	
	\bibitem{Thiemann_2001}{T. Thiemann, Introduction to modern canonical quantum general relativity, arXiv:gr-qc/0110034  (2001).}
	
	\bibitem{Sen_1982}{A. Sen, Gravity as a spin system, Phys. Lett. B \textbf{119}, 89-91 (1982).}
	
	\bibitem{Ashtekar_1986}{A. Ashtekar, new variables for classical and quantum gravity, Phys. Rev. Lett. \textbf{57}, 2244 (1986).}
	
	\bibitem{Barbero_1995}{J. Fernando Barbero, Real Ashtekar variables for Lorentzian signature space-times, Phys. Rev. D \textbf{51}, 5507 (1995).}

	\bibitem{Wilson_1974}{K. G. Wilson, Confinement of quarks, Phys. Rev. D \textbf{10}, 2445 (1974).}
	
	
	\bibitem{Bojowald_2001}{M. Bojowald, Absence of a singularity in loop quantum cosmology, Phys. Rev. Lett. \textbf{86}, 5227 (2001).}
	
	\bibitem{Gambini_2013}{R. Gambini and J. Pullin, Loop quantization of the Schwarzschild black hole, Phys. Rev. Lett. \textbf{110}, 211301 (2013).}
	
	\bibitem{Ashtekar_1998}{A. Ashtekar, J. Baez, A. Corichi, and K. Krasnov, Quantum geometry and black hole entropy, Phys. Rev. Lett. \textbf{80}, 904 (1998).}
	
	\bibitem{Bojowald_review}{M. Bojowald, Loop quantum cosmology, Living Rev. Relativ. \textbf{8}, 11 (2005).}
	
	\bibitem{Ashtekar_2006}{A. Ashtekar, T. Pawlowski, and P. Singh, Quantum nature of the big bang: Improved dynamics, Phys. Rev. D \textbf{74}, 084003 (2006).}
	
	\bibitem{Vakili_2018}{B. Vakili, Classical polymerization of the Schwarzschild metric, Adv. High Energy Phys. \textbf{2018}, 3610543 (2018).}
	
	\bibitem{Bodendorfer_2019}{N. Bodendorfer, Effective quantum extended spacetime of polymer Schwarzschild black hole, Class. Quantum Grav. \textbf{36}, 195015 (2019).}
	
	\bibitem{Achour_2018}{J. B. Achour, F. Lamy, H. Liu, and K. Noui, Polymer Schwarzschild black hole: An effective
		metric, EPL \textbf{123}, 20006 (2018).}
	
	\bibitem{Tibrewala_2012}{R. Tibrewala, Spherically symmetric Einstein–Maxwell theory and
		loop quantum gravity corrections, Class. Quantum Grav. \textbf{29}, 235012 (2012).}
	
	\bibitem{Bardaji_2022a}{A. Alonso-Bardají, An effective model for the quantum Schwarzschild black hole, Phys. Lett. B \textbf{829}, 137075 (2022).}
	
	\bibitem{Bardaji_2022b}{A. Alonso-Bardají, D. Brizuela, and R. Vera, A nonsingular spherically symmetric black-hole model with holonomy corrections, Phys. Rev. D \textbf{106}, 024035 (2022)}
	
	\bibitem{Gambini_2008}{M. Campiglia, R. Gambini, and J. Pullin, Loop quantization of spherically symmetric midi-superspaces, Class. Quantum Grav. \textbf{24} 3649 (2007).}
	
	\bibitem{Bojowald_2005}{M. Bojowald, Loop quantization of spherically symmetric midi-superspaces, Class. Quantum Grav. \textbf{23}, 2129 (2005).}
	
	\bibitem{Dirac}{P. A. M. Dirac, \textit{Lecture in quantum mechanics} (Dover, New York, 2001).}
	
	\bibitem{Bardaji_2021}{A. Alonso-Bardaji, and D. Brizuela, Anomaly-free deformations of spherical general relativity coupled to matter, Phys. Rev. D \textbf{104}, 084064 (2021).}
	
	\bibitem{Beig_1976}{R. Beig, Arnowitt-Deser-Misner energy and $g_{00}$, Phys. Lett. A \textbf{53}, 153 (1976).}
	
	\bibitem{Hayward_1996}{S. A. Hayward, Gravitational energy in spherical symmetry, Phys. Rev. D \textbf{53}, 1938–1949 (1996).}
	
	\bibitem{Schutz}{B. F. Schutz, and C. M. Will, Black hole normal modes: A semianalytic approach,Astrophys.
		J. \textbf{291}, L33 (1985)}
	
	\bibitem{Iyer_1987}{S. Iyer, and C. M. Will, Black-hole normal modes: A WKB approach. I. Foundations and application of a higher-order WKB analysis of potential-barrier scattering . Phys. Rev. D \textbf{35}, 3621 (1987).}
	
	\bibitem{Leaver_1985}{E. Leaver, An Analytic Representation for the quasi-normal modes of Kerr black holes, Proc. R. Soc. A \textbf{402}, 285 (1985).}
	
	\bibitem{Slavyanov}{S. Yu. Slavyanov, and W. Lay, \textit{Special functions - A unified theory based on singularities}
		(Oxford Univertity Press, New York, 2000).}
	
	\bibitem{Leaver_1990}{E.W. Leaver, Quasinormal modes of Reissner-Nordström black holes, Phys. Rev. D \textbf{41}, 2986 (1990).}
	
	\bibitem{Berti_2007}{E. Berti, V. Cardoso, J. A. González, and U. Sperhake, Mining information from binary black hole mergers: A comparison of estimation methods for complex exponentials in noise, Phys. Rev. D \textbf{75}, 124017 (2007).}
	
	\bibitem{Berti_2003}{E. Berti, and K. D. Kokkotas, Asymptotic quasinormal modes of Reissner-Nordström and Kerr black holes, Phys. Rev. D \textbf{68}, 044027 (2003).}
	
	\bibitem{Nollert_1993} H. P. Nollert, Quasinormal modes of Schwarzschild black holes: The determination of quasinormal frequencies with very large imaginary parts, Phys. Rev. D \textbf{47}, 5253 (1993).
	
	\bibitem{Motl_2003}{L. Motl, An analytical computation of asymptotic Schwarzschild quasinormal frequencies, Adv. Theor. Math. Phys. \textbf{6}, 1135 (2003).}
	
	\bibitem{Chandrasekar_1984}{S. Chandrasekhar, On algebraically special perturbations of black holes, Proc. R. Soc. Lond. A \textbf{392}, 1 (1984).}
	
	\bibitem{MN1}{G. Fu, D. Zhang, P. Liu, X. M. Kuang and J. P. Wu, Peculiar properties in quasi-normal spectra from loop quantum gravity effect, arXiv:2301.08421.}

	
	
	
	
	
	
	
	
\end{thebibliography}
\end{document}